\documentclass[reprint,amssymb, amsmath, aps, superscriptaddress, showpacs, footinbib, prl]{revtex4-1}
\pdfoutput=1
\usepackage{graphicx}
\usepackage{color}

\newcommand{\eff}{\text{eff}}
\newcommand{\AFM}{\text{AFM}}

\newcommand{\eps}{\varepsilon}

\newcommand{\Qv}{\mathbf Q}
\newcommand{\dv}{\mathbf d}

\begin{document}

\title{Magnetic frustration in BaCuSi$_2$O$_6$ released}

\author{Vladimir V. Mazurenko}
\email{mvv@dpt.ustu.ru}
\affiliation{Theoretical Physics and Applied Mathematics Department, Ural Federal University, 620002 Ekaterinburg, Russia}

\author{Maria V. Valentyuk}
\affiliation{Theoretical Physics and Applied Mathematics Department, Ural Federal University, 620002 Ekaterinburg, Russia}
\affiliation{Institute of Theoretical Physics, University of Hamburg, Jungiusstra{\ss}e 9, 20355 Hamburg, Germany}

\author{Raivo Stern}
\email{raivo.stern@kbfi.ee}
\affiliation{National Institute of Chemical Physics and Biophysics, 12618 Tallinn, Estonia}

\author{Alexander A. Tsirlin}
\email{altsirlin@gmail.com}
\affiliation{National Institute of Chemical Physics and Biophysics, 12618 Tallinn, Estonia}


\begin{abstract}
Han Purple (BaCuSi$_2$O$_6$) is not only an ancient pigment, but also a valuable model material for studying Bose-Einstein condensation (BEC) of magnons in high magnetic fields. Using precise low-temperature structural data and extensive density-functional calculations, we elucidate magnetic couplings in this compound. The resulting magnetic model comprises two types of nonequivalent spin dimers, in excellent agreement with the $^{63,65}$Cu nuclear magnetic resonance data. We further argue that leading interdimer couplings connect the upper site of one dimer to the bottom site of the contiguous dimer, and not the upper--to-upper and bottom--to--bottom sites, as assumed previously. This finding is verified by inelastic neutron scattering data and implies the lack of magnetic frustration in BaCuSi$_2$O$_6$, thus challenging existing theories of the magnon BEC in this compound.
\end{abstract}

\maketitle
Bose-Einstein condensation (BEC), one of the basic phenomena in quantum physics, has long remained elusive in the experiments until the condensation of bosons in ultracold atomic gases has been observed~\cite{cornell1995,*ketterle1995}. More recently, gapped quantum magnets opened another direction in the experimental studies of the BEC~\cite{giamarchi2008}. Here, individual electronic spins form dimers that can be either in a singlet ($S=0$) or in a triplet ($S=1$) state. The singlet ground state leads to a quantum spin liquid with gapped magnetic excitations, as typical for dimerized spin-$\frac12$ magnets. External magnetic field pushes the triplet state (effective boson) down in energy, so that it eventually becomes populated. Above a certain critical field $H_{c1}$, the concentration of triplets (the chemical potential of bosons) departs from zero, and the system undergoes a BEC transition that manifests itself by a field-induced magnetic ordering observed in thermodynamic measurements~\cite{nikuni2000,*oosawa2001,jaime2004}, neutron scattering~\cite{ruegg2003,*ruegg2008}, and nuclear magnetic resonance (NMR)~\cite{vyaselev2004,kraemer2007} experiments.

The BEC of magnons has been observed and extensively investigated in several model magnetic materials, including BaCuSi$_2$O$_6$~\cite{jaime2004,sebastian2005,sebastian2006,zvyagin2006,kraemer2007,kraemer2013} and TlCuCl$_3$~\cite{nikuni2000,*oosawa2001,vyaselev2004,ruegg2003,*ruegg2008}. In contrast to cold atomic gases, where the spatial arrangement of bosons and their interactions are determined by the external potential, bosonic systems in quantum magnets are to a large extent pre-defined by particular crystal structures and ensuing electronic interactions. For example, in BaCuSi$_2$O$_6$ an unconventional critical exponent reminiscent of a two-dimensional (2D) behavior in an outwardly three-dimensional (3D) spin system has been ascribed to a peculiar pattern of frustrated (competing) interactions between the magnetic layers~\cite{sebastian2006,batista2007}. The magnetic frustration decouples individual layers, thus leading to a dimensional reduction at the quantum critical point at $H_{c1}$, where the spin-liquid phase borders the long-range-ordered BEC phase~\cite{sebastian2006}. 

Later studies of BaCuSi$_2$O$_6$ amended this interesting picture by reporting a low-temperature structural distortion that splits the uniform spin lattice of BaCuSi$_2$O$_6$ into magnetic layers of two different types~\cite{ruegg2007,kraemer2007,sheptyakov2012}. However, all experimental~\cite{sebastian2006,kraemer2013} and theoretical~\cite{batista2007,*schmalian2008,vojta2007,*vojta2007-2,laflorencie2009,*laflorencie2011} studies available so far consider magnetic frustration as an integral part of BaCuSi$_2$O$_6$. Here, we challenge this well-established paradigm by evaluating individual magnetic couplings in BaCuSi$_2$O$_6$ from density-functional (DFT) calculations and re-analyzing the neutron-scattering data. We find that BaCuSi$_2$O$_6$ is essentially a non-frustrated magnet, and suggest that the mechanism of the magnon BEC in this compound should be reconsidered.    

\begin{figure}
\includegraphics{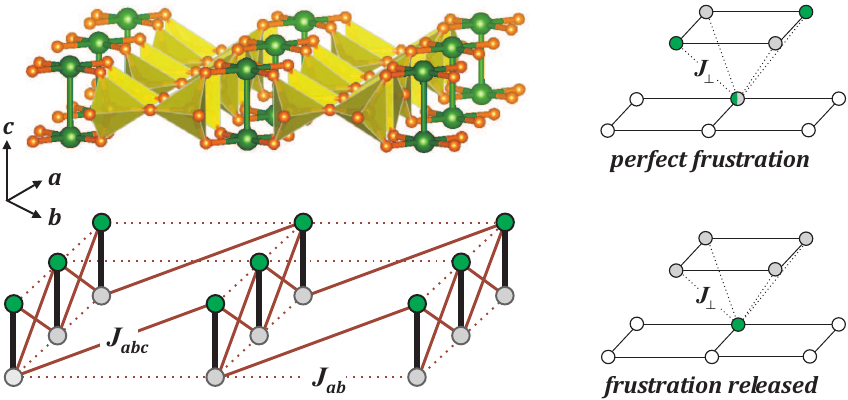}
\caption{\label{fig:structure}
(Color online) Left panel: crystal structure of the magnetic bilayer and the relevant magnetic model with the FM in-plane order driven by the AFM interdimer cooupling $J_{abc}\gg J_{ab}$. Green (dark) and gray (light) circles denote different spin directions. Right panel: different regimes of interlayer interactions depending on the in-plane magnetic order. The AFM in-plane order leads to a perfect frustration (top). The FM in-plane order lifts the frustration (bottom). Crystallographic plots are done using the \texttt{VESTA} software~\cite{vesta}.
}
\end{figure}

BaCuSi$_2$O$_6$ is colloquially known as Han Purple, the pigment used in ancient China~\cite{berke2007}. Its room-temperature crystal structure, although unknown to the original Chinese users, is tetragonal (space group $I4_1/acd$) and features CuO$_4$ square plaquettes connected via Si$_4$O$_{10}$ ring units of corner-sharing SiO$_4$ tetrahedra (Fig.~\ref{fig:structure})~\cite{sparta2004}. The plaquettes are linked in such a way that two Cu atoms are separated by 2.75~\r A only, forming a well-defined structural and magnetic dimer. The dimers are arranged into slabs, thus forming magnetic bilayers (Fig.~\ref{fig:structure}). It is commonly believed that the in-plane order is antiferromagnetic (AFM), driven by the coupling $J_{ab}$. 

The bilayers are interleaved by Ba$^{2+}$ cations. The stacking of the bilayers is such that each spin-$\frac12$ Cu$^{2+}$ ion interacts with four Cu$^{2+}$ ions of the neighboring layer, two of them having one spin direction and the two other having an opposite spin direction, because the in-plain order is, presumably, AFM (Fig.~\ref{fig:structure}, top right). This way, the interlayer couplings $J_{\perp}$ are perfectly frustrated, no matter whether $J_{\perp}$ is ferromagnetic (FM) or AFM. While these coupling are very weak, likely below 1~K~\cite{jaime2004,ruegg2007}, their allegedly frustrated nature prevents the system from a 3D ordering. This crucial microscopic feature underlies the idea of the dimensional reduction at the QCP. However, the frustrated nature of $J_{\perp}$ is invalidated by our detailed microscopic analysis of BaCuSi$_2$O$_6$ reported below.

In the following, we evaluate individual magnetic couplings in BaCuSi$_2$O$_6$ using DFT band-structure calculations performed in the \texttt{FPLO}~\cite{fplo} code. From a tight-binding analysis of the band structure calculated on the level of local density approximation (LDA), we obtain hopping integrals $t_i$ that are related to AFM exchange integrals as $J_i^{\AFM}=4t_i^2/U_{\eff}$, where $U_{\eff}$ is an effective on-site Coulomb repulsion in the Cu $3d$ shell. Alternatively, we estimate exchange couplings $J_i$ as energy differences between collinear FM and AFM spin configurations calculated within the LSDA+$U$ approach, where the Hubbard $U$ parameter accounts for electronic correlation in a mean-field fashion~\cite{tsirlin2010,*tsirlin2012}. We have cross-checked the above \texttt{FPLO} results using total-energy calculations in the \texttt{VASP} code~\cite{vasp1,*vasp2} and the Lichtenstein exchange integral procedure (LEIP)~\cite{leip} implemented in \texttt{TB-LMTO-ASA}~\cite{lmto}. All these approaches provide the consistent microscopic scenario of BaCuSi$_2$O$_6$~\footnote{Our LSDA+$U$ calculations yield the band gap of about 3.3~eV and the magnetic moment of 0.82~$\mu_B$ on Cu atoms at $U_d=6.5$~eV and $J_d=1$~eV.}.

We start with the room-temperature $I4_1/acd$ structure of BaCuSi$_2$O$_6$ (Table~\ref{tab:exchange}). Here, both $t_i$ and $J_i$ support the overall model of spin dimers forming the bilayers. However, we find that the leading interdimer coupling within the bilayer is clearly $J_{abc}$ and not $J_{ab}$. The upper site of one dimer is coupled to the bottom site of the neighboring dimer, thus leading to the FM in-plane order. This FM order lifts the frustration of $J_{\perp}$ (Fig.~\ref{fig:structure}). 

\begin{figure}
\includegraphics[width=8cm]{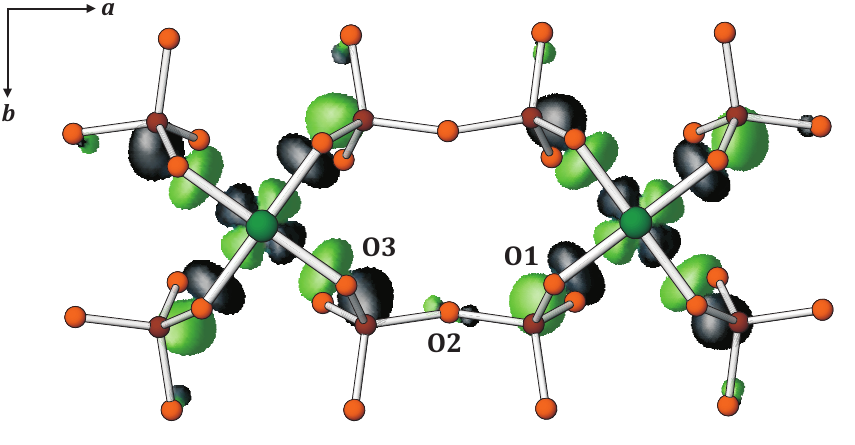}
\caption{\label{fig:wannier}
(Color online) Cu-based Wannier functions showing the mechanism of the Cu--O1--O2--O3--Cu superexchange in BaCuSi$_2$O$_6$.
}
\end{figure}
The unexpected $J_{abc}\gg J_{ab}$ coupling regime can be rationalized by considering individual atomic orbitals that contribute to the electron hopping and, therefore, to the superexchange process. Fig.~\ref{fig:wannier} shows Wannier functions based on the Cu $3d_{x^2-y^2}$ orbitals. Each Wannier function comprises $p\sigma$ contributions of four nearest-neighbor oxygen atoms O1 and O3 (about 10\,\% each) and, additionally, four smaller contributions of second-neighbor oxygens O2 (about 0.5\,\% each). These second-neighbor contributions are not unusual and largely determine the superexchange in Cu$^{2+}$ magnets~\cite{tsirlin2010,*tsirlin2012}. In BaCuSi$_2$O$_6$, the superexchange follows the Cu--O1--O2--O3--Cu pathway and, therefore, crucially depends on the O1--O2--O3 angle, which is $95.5^{\circ}$ for $J_{ab}$ and $156.1^{\circ}$ for $J_{abc}$. Therefore, $J_{abc}$ should be the leading interdimer coupling because of the more favorable orbital overlap according to the conventional Goodenough-Kanamori-Anderson rules~\cite{gka}.

\begin{figure}
\includegraphics{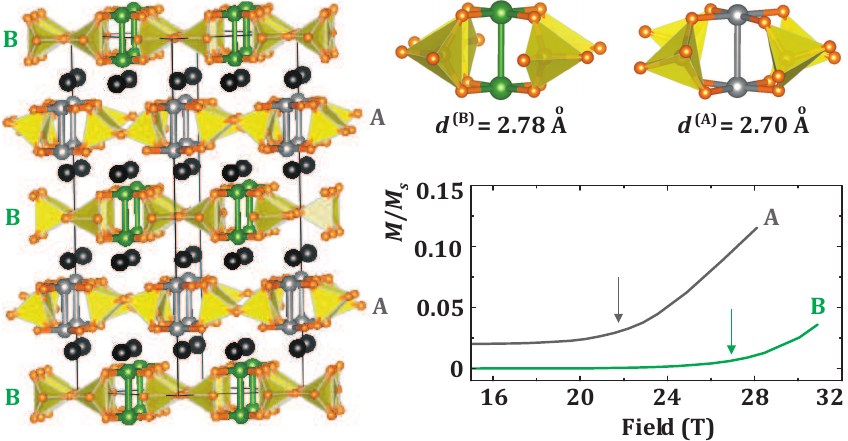}
\caption{\label{fig:mvsh}
(Color online) Low-temperature orthorhombic structure of BaCuSi$_2$O$_6$ with two inequivalent dimers A and B. Bottom right part shows the simulated magnetization curves of individual bilayers at $T/J=0.05$. The arrows mark the critical fields of $H_{c1}^{(A)}\simeq 22$~T and $H_{c1}^{(B)}\simeq 27$~T, where the relevant spin gaps are closed.
}
\end{figure}

\begin{table}
\caption{\label{tab:exchange}
Magnetic couplings in the high-temperature and low-temperature phases of BaCuSi$_2$O$_6$: Cu--Cu distances $d$ (in~\r A), transfer integrals $t_i$ (in~meV), and exchange integrals $J_i$ (in~K). Note that the $t_i$ values represent AFM contributions to the exchange, according to $J_i^{\AFM}=4t_i^2/U_{\eff}$, where $U_{\eff}$ is the effective on-site Coulomb repulsion. Full exchange couplings $J_i$ are obtained from LSDA+$U$ with $U_d=6.5$~eV, $J_d=1$~eV and the around-mean-field double-counting correction~\cite{tsirlin2010,*tsirlin2012}.
}
\begin{ruledtabular}
\begin{tabular}{crrrrrrrr}
     & \multicolumn{2}{c}{$J$}   & \multicolumn{2}{c}{$J_{ab}$} & \multicolumn{2}{c}{$J_{abc}$} & \multicolumn{2}{c}{$J_{\perp}$} \\
 $d$ & \multicolumn{2}{c}{2.75}  & \multicolumn{2}{c}{7.08}  & \multicolumn{2}{c}{7.59}     & \multicolumn{2}{c}{5.77}  \\
 $t$ & \multicolumn{2}{c}{$-93$} & \multicolumn{2}{c}{$-2$}  & \multicolumn{2}{c}{36}       & \multicolumn{2}{c}{$-5$}  \\
 $J$ & \multicolumn{2}{c}{53}    & \multicolumn{2}{c}{$-0.2$} & \multicolumn{2}{c}{7.9}     & \multicolumn{2}{c}{0.4}   \\\hline
     & $J^A$  & $J^B$ & $J_{ab}^A$ & $J_{ab}^B$ & $J_{abc}^A$ & $J_{abc}^B$ & $J_{\perp}^A$ & $J_{\perp}^B$ \\
 $d$ &  2.70  &  2.78 &   7.04    &   7.04    &   7.54     &    7.57    &   5.73  &   5.72  \\
 $t$ & $-88$  & $-105$ &   2      &   $-10$   &    33      &    38      &   $-4$  &   $-5$  \\
 $J$ &   58   &   68  &   $-0.2$  &   $-0.2$  &    7.2     &    8.0     &    0.2  &    0.2  \\
\end{tabular}
\end{ruledtabular}
\end{table}
Let us now consider the changes in the magnetic model upon the transition to the low-temperature orthorhombic $Ibam$ structure around 100~K~\cite{sheptyakov2012}. Here, tiny alterations in the mutual arrangement of the CuO$_4$ and SiO$_4$ units render two neighboring bilayers inequivalent (Fig.~\ref{fig:mvsh}). These bilayers labeled A and B feature different intradimer couplings, as seen by the inelastic neutron scattering (INS)~\cite{ruegg2007} and NMR~\cite{kraemer2007}. Indeed, we find two types of the intradimer exchange $J^A$ and $J^B$ (Table~\ref{tab:exchange}). All other couplings also split into two, but only $J^A$ and $J^B$ reveal a sizable difference. The interdimer exchange remains largely unchanged, so that the $J_{abc}\gg J_{ab}$ regime persists at low temperatures. 

The structural difference between the A and B dimers is well seen in the Cu--Cu distances that are 2.70~\r A and 2.78~\r A, respectively. Surprisingly, we find that the AFM exchange is stronger for the longer Cu--Cu distance ($J^B$) and weaker for the shorter Cu--Cu distance ($J^A$), although a naive picture of the direct $d-d$ overlap should be exactly the opposite. This unexpected result should be traced back to the nature of interacting orbitals visualized by the Cu-based Wannier functions in  Fig.~\ref{fig:wannier}. The relevant orbitals lie in the CuO$_4$ plane, thus making the direct $d-d$ exchange impossible. The intradimer coupling follows the long Cu--O1--O2--O3--Cu superexchange pathway instead, similar to the case of $J_{abc}$. Therefore, the intradimer exchange $J^{A,B}$ lacks any simple relation to the Cu--Cu distance $d^{A,B}$. 

The emergence of a weaker coupling in the shorter Cu--Cu dimer is independently confirmed by the $^{63,65}$Cu NMR experiments~\cite{kraemer2007}. Here, the A and B dimers are distinguished according to their different spin gaps and different quadrupolar frequencies on the respective Cu sites. The B dimer with the larger spin gap (i.e., the stronger intradimer coupling) shows the smaller EFG, and the other way around. To verify that the strongly coupled B dimer is indeed the one with the longer Cu--Cu distance, we calculated the quadrupolar frequencies of 33.8~MHz and 30.5~MHz for the A and B dimers, respectively. These values should be compared with 14.85~MHz and 14.14~MHz from the NMR experiment~\cite{kraemer2007}. While the absolute values of the quadrupolar frequencies are substantially overestimated and reflect well-known shortcomings of DFT in evaluating subtle features of charge distribution and electric field gradients, the qualitative effect of the smaller frequency on the stronger dimer is well reproduced. Therefore, the DFT and NMR results are in good agreement.

\begin{figure}
\includegraphics{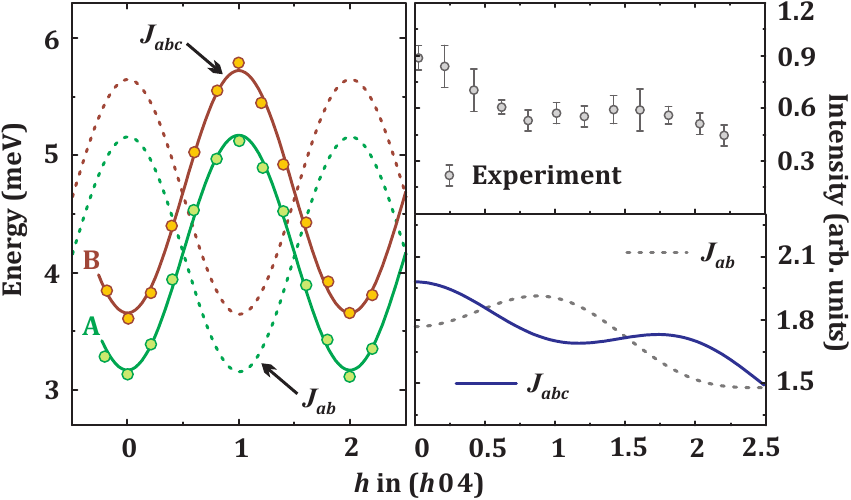}
\caption{\label{fig:neutron}
(Color online) Left panel: dispersions of the triplet bands corresponding to the A and B bilayers in BaCuSi$_2$O$_6$. Solid and dashed lines are drawn according to Eq.~\eqref{eq:dispersion} for $J_{abc}\gg J_{ab}$ and $J_{ab}\gg J_{abc}$ regimes, respectively. Right panel: momentum dependence of the INS intensity (top) and the predictions of Eq.~\eqref{eq:int} for the $J_{abc}\gg J_{ab}$ (solid line) and $J_{ab}\gg J_{abc}$ (dashed line) scenarios (bottom). Experimental data from Ref.~\onlinecite{ruegg2007} are taken along the $(h04)$ direction of the reciprocal space. 
}
\end{figure}
We can further make a quantitative comparison to the experiment. INS studies~\cite{ruegg2007} yield $J^A\simeq 50$~K and $J^B\simeq 55$~K in remarkable agreement with our DFT estimates listed in Table~\ref{tab:exchange}. Regarding the interdimer exchange, Ref.~\onlinecite{ruegg2007} reports $J_{ab}\simeq 6$~K (AFM), whereas an earlier work~\cite{zheludev1997} puts forward the FM coupling $J_{ab}\simeq -2.2$~K. This latter result supports our key finding of $J_{abc}$ as the leading interdimer coupling. While $J_{abc}$ is AFM, it leads to the FM order in the $ab$ plane (Fig.~\ref{fig:structure}) that can be seen as an effective FM coupling $J_{ab}$ probed by the neutron scattering. In fact, the same conclusion is inferred from the data of Ref.~\onlinecite{ruegg2007}. The dispersion of the triplet band can be written as follows:
\begin{equation}
  \eps=J+(J_{ab}-J_{abc})[\cos(\pi h+\pi k)+\cos(\pi h-\pi k)],
\label{eq:dispersion}\end{equation}
which, considering the experimental dispersion~\cite{ruegg2007,zheludev1997}, implies a positive $J_{abc}$ (or negative $J_{ab}$, as in Ref.~\onlinecite{zheludev1997}). A positive $J_{ab}\gg J_{abc}$ would manifest itself by energy minima at odd $h$, in sharp contrast to the experimental dispersion showing the minima at even $h$, including $h=0$ (Fig.~\ref{fig:neutron}, left). The different positions of the minima result from the fact that $J_{ab}$ and $J_{abc}$ lead to different ordering patterns in the $ab$ plane (Fig.~\ref{fig:structure}). The order established by $J_{ab}$ is AFM, whereas $J_{abc}$ would favor the FM in-plane order. 

To calculate the scattering intensity, we used the single mode approximation~\cite{stone2007}. In our case, $I(\Qv,\omega)=I(\Qv)\delta(\omega-\omega_{\Qv})$, where $\omega_{\Qv}$ is determined by Eq.~\eqref{eq:dispersion}. Therefore, we evaluate the intensities using the equation:
\begin{equation}
  I(\Qv)\sim -|F(Q)|^2\sum_{\dv}A_{\dv}(1-\cos(\Qv\dv)),
\label{eq:int}\end{equation}
where $\dv$ are vectors connecting the Cu sites, $A_{\dv}=J_{\dv}\langle S_0S_{\dv}\rangle$, and for the sake of simplicity we consider the tetragonal $I4_1/acd$ crystal structure, because the effect of $J_{abc}\gg J_{ab}$ pertains to both polymorphs. The calculation for the relevant $\Qv$ values shows that the model with $J_{abc}\gg J_{ab}$ reproduces the minimum of the intensity at $h\simeq 1$ (Fig.~\ref{fig:neutron}, right). In contrast, the model with $J_{abc}\ll J_{ab}$ will lead to a maximum at $h=1$ in apparent contradiction to the experiment~\footnote{Here, we used $A=1$ and $A_{ab}=0.03$ in order to match the experimental data.}. Therefore, the INS data strongly support the proposed microscopic magnetic model with $J_{abc}\gg J_{ab}$.

Our computed exchange couplings also reproduce the critical fields $H_{c1}$ of individual bilayers, as shown by the magnetization isotherms calculated with quantum Monte-Carlo (QMC) algorithm implemented in the \texttt{ALPS} code~\cite{alps}. The departure of the magnetization from zero signifies the closing of the spin gap. We find $H_{c1}^A\simeq 22$~T and $H_{c1}^B\simeq 27$~T to be compared with $H_{c1}\simeq 23.4$~T obtained experimentally~\cite{kraemer2007}. This critical field corresponds to the BEC transition in the A bilayer with the smaller spin gap. The inter-bilayer coupling generates triplets (bosons) in the B bilayer as well, even though its spin gap is not yet closed at $H_{c1}$~\cite{kraemer2007}. This effect, which is arguably the most peculiar feature of BaCuSi$_2$O$_6$ relating to the dimensional reduction at the QCP, has been a matter of substantial theoretical interest~\cite{vojta2007,*vojta2007-2,laflorencie2009,*laflorencie2011}. It is presently understood as a consequence of frustrated inter-bilayer couplings $J_{\perp}$ that, however, should be only partially frustrated in order to explain the experimental data available so far~\cite{kraemer2013}. Our microscopic results challenge this arduously reached understanding, because the FM in-plane order, as directly inferred from DFT and from the neutron data, \emph{leaves no room for the frustration} in BaCuSi$_2$O$_6$. 

Apparently, further theoretical and experimental work will be required to reconcile experimental observations regarding the BEC transition in BaCuSi$_2$O$_6$. The diagonal interdimer coupling $J_{abc}$ and the ensuing FM in-plane order should be considered as basic ingredients of the minimum magnetic model. The inter-bilayer coupling is about 0.2~K only, but it does not prevent and rather facilitates the 3D magnetic order in this compound. Further, and more subtle ingredients, may be weak incommensurate structural modulations that were mentioned in Ref.~\onlinecite{samulon2006} but never observed in the subsequent neutron-diffraction study~\cite{sheptyakov2012}. We conclude that BaCuSi$_2$O$_6$, a seemingly well-known model BEC compound, still keeps a lot of puzzles that require further investigation.

\acknowledgments
The work of VVM and MVV was supported by the grant program of President of Russian Federation MK-5565.2013.2, the contracts of the Ministry of Education and Science of Russia N 14.A18.21.0076 and 14.A18.21.0889. AT acknowledges financial support from the ESF via the Mobilitas program (grant MTT77). RS was funded by the Estonian Research Council (grants ETF8440 and PUT210). We acknowledge essential discussions with Cristian Batista and Christian R\"uegg as well as fruitful communication with F. Mila, V. I. Anisimov, and A. I. Lichtenstein.

%

\end{document}